# Evidence of Spin Frustration in Vanadium Diselenide Monolayer Magnet

*Ping Kwan Johnny Wong, Wen Zhang, Fabio Bussolotti, Xinmao Yin, Tun Seng Herng, Lei Zhang, Yu Li Huang, Giovanni Vinai, Sridevi Krishnamurthi, Danil W. Bukhvalov, Yu Jie Zheng, Rebekah Chua, Alpha T. N'Diaye, Simon A. Morton, Chao-Yao Yang, Kui-Hon Ou Yang, Piero Torelli, Wei Chen, Kuan Eng Johnson Goh, Jun Ding, Minn-Tsong Lin, Geert Brocks, Michel P. de Jong,\* Antonio H. Castro Neto,\* and Andrew Thye Shen Wee\**

Dr. P. K. J. Wong, L. Zhang, Dr. W. Chen, Prof. A. H. Castro Neto, Prof. A. T. S. Wee
Centre for Advanced 2D Materials (CA2DM) and Graphene Research Centre (GRC),
National University of Singapore, 6 Science Drive 2, Singapore 117546, Singapore
E-mail: phycastr@nus.edu.sg; phyweets@nus.edu.sg

Dr. W. Zhang, Dr. X. M. Yin, L. Zhang, Dr. Y. L. Huang, Dr. Y. J. Zheng, R. Chua, Dr. W. Chen, Dr. K. E. J. Goh, Prof. A. H. Castro Neto, Prof. A. T. S. Wee
Department of Physics, National University of Singapore, 2 Science Drive 3, Singapore 117542, Singapore

Dr. F. Bussolotti, Dr. K. E. J. Goh
Institute of Materials Research and Engineering (IMRE), Agency for Science, Technology and Research (A*Star), 2 Fusionopolis Way, Innovis, Singapore 138634, Singapore

Dr. T. S. Herng, Prof. J. Ding
Department of Materials Science and Engineering, National University of Singapore, 9 Engineering Drive 1, Singapore 117575, Singapore

Dr. G. Vinai, Dr. P. Torelli
Instituto Officina dei Materiali (IOM)-CNR, Laboratorio TASC, Area Science Park, S.S. Km 163.5, Trieste I-34149, Italy

S. Krishnamurthi, Prof. G. Brocks
Computational Materials Science, Faculty of Science and Technology and MESA+ Institute for Nanotechnology, University of Twente, P.O. Box 217, 7500 AE Enschede, The Netherlands

Prof. D. W. Bukhvalov
College of Science, Institute of Materials Physics and Chemistry, Nanjing Forestry University, Nanjing 210037, P. R. China
Institute of Physics and Technology, Ural Federal University, Mira Street 19, 620002 Yekaterinburg, Russia

Dr. A. T. N'Diaye, Dr. S. A. Morton
Advanced Light Source (ALS), Lawrence Berkeley National Laboratory, Berkeley CA94720, USA

Dr. C.-Y. Yang, K.-H. Ou Yang, Prof. M.-T. Lin
Department of Physics, National Taiwan University, Taipei 10617, Taiwan

Dr. W. Chen
Department of Chemistry, National University of Singapore, 2 Science Drive 3, Singapore 117542, Singapore

Dr. M. P. de Jong
NanoElectronics Group, MESA+ Institute for Nanotechnology, University of Twente, P.O. Box 217, 7500 AE, Enschede, The Netherlands
E-mail: m.p.dejong@utwente.nl

**Monolayer $VSe_2$, featuring both charge density wave and magnetism phenomena, represents a unique van der Waals magnet in the family of metallic two-dimensional transition-metal dichalcogenides (2D-TMDs). Herein, by means of *in-situ* microscopic and spectroscopic techniques, including scanning tunneling microscopy/spectroscopy, synchrotron X-ray and angle-resolved photoemission, and X-ray absorption, direct spectroscopic signatures are established, that identify the metallic 1$T$-phase and vanadium $3d^1$ electronic configuration in monolayer $VSe_2$ grown on graphite by molecular-beam epitaxy. Element-specific X-ray magnetic circular dichroism, complemented with magnetic susceptibility measurements, further reveals monolayer $VSe_2$ as a frustrated magnet, with its spins exhibiting subtle correlations, albeit in the absence of a long-range magnetic order down to 2 K and up to a 7 T magnetic field. This observation is attributed to the relative stability of the ferromagnetic and antiferromagnetic ground states, arising from its atomic-scale structural features, such as rotational disorders and edges. The results of this study extend the current understanding of metallic 2D-TMDs in the search for exotic low-dimensional quantum phenomena, and stimulate further theoretical and experimental studies on van der Waals monolayer magnets.**

Endowed by the many possible combinations of their constituting elements, two-dimensional transition-metal dichalcogenides (2D-TMDs) can exhibit a multitude of exotic properties.[1, 2] For the much-studied group-VI 2D semiconductors, these include coupled spin and valley degrees of freedom,[3] an indirect-to-direct bandgap transition,[4, 5] valley polarization,[6, 7] long-lived electron spin in the nanosecond time regime,[8] and so on. There is increasing interest in metallic 2D-TMDs in which a rich variety of intrinsic correlated electronic phases can be experimentally accessed in the monolayer limit.[9-12] These electronic phases can be exploited as potential building blocks for functional van der Waals heterostructures,[13, 14] as well as for other electronic,[15, 16] sensing,[17] or catalytic[18, 19] applications.

As an example of metallic 2D-TMDs, VSe$_2$ can theoretically exhibit two polymorphs, *i.e.* the 1$T$ and 2$H$ structures, but only the former is known to exist.[20-22] The metallic character of VSe$_2$ stems from its 3$d^1$ electron configuration, which also invokes charge-density wave (CDW) order and magnetism. Recent scanning tunneling microscopy (STM) and angle-resolved photoemission (ARPES) studies have confirmed an enhancement of the CDW order in monolayer VSe$_2$ grown by molecular-beam epitaxy (MBE).[23-26] However, the ferromagnetic state, predicted by first-principles calculations, remains controversial, as conflicting results have been reported.[21, 23-25, 27, 28] Theoretically, the magnetic moment in monolayer VSe$_2$, mainly arising from the vanadium ions with a value of ~0.6 $\mu_B$ per V ions,[21, 22] should be much smaller than those in 3$d$ ferromagnets, such as Fe (~2.2 $\mu_B$) or Co (~1.8 $\mu_B$). Experimentally, a huge value of almost an order larger than Fe and Co has been reported, however.[23] Such inconsistency has thus cast doubt on the sole use of bulk magnetic tools for characterizing monolayer VSe$_2$ and other 2D magnets, where its capability to extract a weak signal from an overwhelmingly larger substrate background signal can potentially lead to misleading results.[29] A similar issue has been encountered previously with studies on diluted magnetic semiconductors, where unintended magnetic contributions from various sources (contaminations, defects, *etc*.) led to observations that were thought to be intrinsic to the materials.[30] Therefore, to avoid any conflicting observations, careful assessments by comparison to other unambiguous techniques would be mandatory.

In this work, we utilize a combination of *in-situ* microscopic and spectroscopic techniques, including STM/STS, synchrotron-based PES, ARPES and X-ray absorption (XAS), to elucidate the metallic 1*T*-phase and $3d^1$ electronic configuration of monolayer VSe$_2$ grown on HOPG by MBE. In stark contrast to previous reports,[23, 24] our element-specific X-ray magnetic circular dichroism (XMCD) measurements finds no salient feature of ferromagnetically-coupled V 3*d* states down to 16 K and up to a 1 T magnetic field. We, instead, observe traits of spin frustration in the monolayer from complementary temperature- and field-dependent susceptibility measurements. Our results are attributed to atomic-scale structural details of the monolayer, which can affect the relative stability of the ferromagnetic and antiferromagnetic ground states. This new finding is expected to open up new and careful searches for exotic low-dimensional quantum phenomena.

The 1*T* structure of VSe$_2$ consists of a hexagonal plane of V atoms sandwiched by two Se atomic planes with in-plane and out-of-plane lattice constants of 3.4 and 6.1 Å (**Figure 1**a).[20] For its monolayer on HOPG, we find similar lattice structure and parameters by STM at 77 K. The line profile (Figure 1b) across the step edge of a VSe$_2$ terrace indicates an apparent height of 7.0 Å, in good agreement with the monolayer height for VSe$_2$. Our atomic-resolution STM measurement (Figure 1c) reveals a (1 × 1) hexagonal structure unit with an in-plane lattice constant of 3.4 ± 0.1 Å, as marked by the line profile in Figure 1d. We also observe two additional ($\sqrt{3} \times 2$) and ($\sqrt{3} \times \sqrt{7}$) superstructures, similarly reported in a recent STM study.[24] These superstructures have been associated with a low-temperature CDW, with enhanced charge ordering arising from an energy gap opening in the monolayer 2D Fermi surface rather than in only the well-nested portions as in the bulk.[25] Our tunneling spectroscopy (STS) measurements identify the presence of such a CDW gap, manifested as a differential conductance dip around the Fermi level. The size of this gap is ~26 meV, comparable with values reported previously.[23-25] It is worthwhile to note that the much larger CDW gap observed in a prior ARPES measurement[26] is related to the fact that such gap was obtained along the M–K direction of the VSe$_2$'s 2D Brillouin zone, which has a more pronounced gap opening (~100 meV) than that along the Γ–K direction (in the order of ~20 meV or smaller). This essentially explains the smaller CDW gap in our case, as the STS mainly probes the local density-of-states around the Γ point.

**Figure 2** summarizes the surface electronic properties of monolayer VSe$_2$ measured by synchrotron-PES. Core-level measurements (Figure 2a) show that the V 2$p$ characteristic peaks of the monolayer are positioned at 513.4 and 520.9 eV in binding energy (BE), and the Se 3$d$ peaks at 53.4 and 54.2 eV. For atomically thin materials, one should consider the role of substrate-induced core-hole screening in defining the measured core-level positions. As a general rule, the more conductive a substrate is, the stronger the screening of the overlayer core-hole. Hence, the overlayer core-levels will be shifted to lower BE. We note that this screening mechanism is rarely discussed in previous studies of 2D-TMDs,[18, 31-34] and explains the lower core-level positions of monolayer VSe$_2$ on HOPG compared to those (V 2$p_{3/2}$ at 514–517 eV) on insulating substrates (SiO$_2$ and mica). However, as we will discuss below, inevitable oxidation of ambient-exposed VSe$_2$ can also influence these core-level positions.

The valence band and work function of monolayer VSe$_2$ are probed by a photon source with an energy of 60 eV. A sharp Fermi edge is observed at the valence band (Figure 2c), originating from the V 3$d$ states. The spectral features at 1–6 eV are assigned to the Se 4$p$ derived bands, while beyond 6 eV lies a substrate-related bump. From the secondary electron cutoff (Figure 2d), we obtain a work function of 5.0 ± 0.1 eV. This value, an average over the whole sample surface, has been calibrated against that of HOPG (Figure S1, Supporting Information) and is comparable with the value experimentally extracted by STM.[35] Given its good match to that of monolayer MoSe$_2$ (5.1 eV), our measured value raises a tantalizing prospect of a low-resistivity p-type metal/semiconductor van der Waals heterostructure.[36]

Air exposure result in newly developed peaks at the higher BE side of the initial V $2p$ core-levels, in addition to a high-BE shoulder in the Se $3d$ spectrum. These new peaks indicate oxide formation, given their similar BE positions to those of $VO_2$.[37] Both the valence band and work function are significantly modified. The initially sharp Fermi edge is seen to lose almost all of its spectral weight after the exposure, accompanied by a work function reduction of 0.25 eV. These are collectively comparable to published results comparing pristine and contaminated $VSe_2$ crystals.[38]

**Figure 3**a shows the $d^1$ ($V^{4+}$) electronic configuration for $1T$-$VSe_2$. Due to the octahedral coordination, the $3d$ degenerate orbitals of the V ions are split into two sets of $e_g$ and $t_{2g}$ states, with an energy separation of 10 $Dq$. The $t_{2g}$ states are at lower energy orbitals than the $e_g$ states and therefore are filled by an unpaired electron. By measuring its V $L_{2,3}$-edge XAS in the total electron yield (TEY) geometry (Figure 3b),[39, 40] we directly verify this $d^1$ configuration in our monolayer. The XAS spectrum in Figure 3c reveals two main peaks (feature a,c) at ~518 and 524 eV, corresponding to the V $L_3$ and $L_2$ absorption edges due to dipole-allowed atomic-like transitions from the spin-orbit split V $2p_{3/2}$ and $2p_{1/2}$ core-levels to the $3d$ unoccupied states. The fine structures (labeled as a,b,d) at the lower energy side of those main peaks are related to the distribution of atomic multiplets. These spectral features are akin to those measured from bulk $1T$-$VS_2$[33] and other $3d^1$ vanadium compounds,[41] thus providing a spectroscopic fingerprint of the $1T$ phase of the monolayer.

In agreement with ref. [25], our V $L_{2,3}$-edge XMCD measurements, performed in an applied magnetic field up to 1 T and at temperatures down to 16 K (lower panel of Figure 3c and Figure S2, Supporting Information), rules out intrinsic ferromagnetism in the monolayer. The measured XMCD values are less than 0.2%, far below the detection sensitivity of the synchrotron facilities used. The monolayer's SQUID $M$–$H$ curves (**Figure 4**a,b) support this finding and further reveal no sign of magnetic hysteresis but a linear behavior that persists down to 2 K (see Figure S3, Supporting Information, for Se-capped HOPG signal). Such a linear behavior is anisotropic in nature, with a higher moment value along the out-of-plane direction than that within the 2D plane. The positive slope for the out-of-plane direction indicates a paramagnetic behavior, whereas the negative slope obtained within the 2D plane suggests a diamagnetic response. The latter is a primary result of the difficulty in separating the much weaker monolayer signal from the more predominant substrate background, a general issue known for other existing 2D magnets.[23, 29]

Magnetic susceptibility ($\chi = M/H$, where $M$ and $H$ are defined as magnetic moment per area and magnetic field strength, respectively) of the monolayer further reveals clear signatures of spin frustration (Figure 4c). One observes an overlap between the field-cooling (FC) and zero-field-cooling (ZFC) curves, indicating an absence of a long-range magnetic order, even down to 2 K and in a 7 T field (upper panel). One also observes a broad maximum for the measured $\chi$, a characteristic feature of low-dimensional magnetic systems with a short-range antiferromagnetic (AF) interaction.[42] A negative Weiss constant $\theta_{cw}$ obtained from the Curie-Weiss (CW) fits to the high-temperature region of $\chi^{-1}$ (lower panel of Figure 4c and Figure S4, Supporting Information) indeed supports an AF origin of the exchange interaction in monolayer VSe$_2$. This interaction is, however, distinguished from that of a conventional AF magnet, because of its field-dependent $\theta_{cw}$, as compared between Figure 4c and Figure S4. For the 7 T field data, we obtain a frustration factor $f = |\theta_{cw}|/T_N \geq 51$, where the AF Néel temperature $T_N$ is below 2 K. The fact that $f$ is much larger than unity[43] gives further evidence of spin frustration in monolayer VSe$_2$. The frustration observed here can have a geometrical origin, given a reasonable reproducibility of the temperature-dependent $\chi$ by the 2D spin-1/2 Heisenberg antiferromagnet on a triangular lattice model[44, 45] (Figure S5, Supporting Information). The essential idea of this model is depicted in Figure 4d, where, on a triangular spin lattice with AF coupling, three neighboring spins cannot be mutually antialigned with each other, thus resulting in fluctuating spins and suppressed AF correlation, in accordance to our observation. It is worth mentioning that, due to the small spins (*i.e.* spin-1/2) and monolayer nature of our VSe$_2$, other contributions, such as quantum frustration and low-dimensionality, may also play a role in the observed frustration, because of their tendencies to destabilize long-range magnetic order.

Given the prior predictions of a more stable ferromagnetic ground state than that of AF in monolayer VSe$_2$,[21, 22, 28] our findings above may appear surprising. However, DFT calculations also predict the 2*H* phase to be slightly more stable than the 1*T* phase, which in view of our ARPES measurements is incorrect (**Figure 5** and Figure S6, Supporting Information). Inclusion of a Hubbard on-site repulsion *U* in the DFT calculations, however, stabilizes both the 1*T* structure, as well as the AF ordering. Nevertheless, the energy difference with ferromagnetic ordering is small (~22 meV, Table S1, Supporting Information), which can be lifted by the presence of structural features, such as misaligned domains and boundaries, vacancies, or edges. These can contribute additional states at the Fermi level, as corroborated by ARPES. Figure 5a shows an ARPES intensity map of monolayer VSe$_2$ measured at 300 K. Near the Fermi level, one observes a weakly dispersive V 3*d* band and a set of degenerate Se 4*p* bands with a strong downward dispersion. These bands are better distinguished in Figure 5b,c. Although these measured bands are azimuthally-averaged results of differently oriented domains (Figure S7, Supporting Information), the good agreement in the overall band structure between our experiment and DFT calculations (Figure S6, Supporting Information) confirms the corresponding 1*T* structure of the monolayer. Our findings are also consistent with recent reports that the intrinsic valence electron structure of monolayer VSe$_2$ is largely preserved on graphitic systems.[25, 26] Upon cooling to 12 K, a CDW transition is evident, shifting the V 3*d* and Se 4*p* bands away from the Fermi level, with a marked increase in their intensities (Figure 5d–f). These bands appear as a double-peak structure in the energy distribution curves (EDCs) in Figure 5g, which is otherwise unresolved at 300 K. The relative spectral changes observed are further analyzed by normalizing the EDC intensities to the Fermi-Dirac (FD) function, in order to highlight the effect of temperature on the spectral broadening. Essentially, the EDCs indicate a leading-edge midpoint shift toward higher BE upon cooling, indicative of the opening of a CDW gap, in agreement with our STS measurements as well as previous studies.[23, 25, 26] However, we note that a full gap has been reported previously,[25] in contrast to our observation of a finite Fermi-edge cut-off, even at 12 K. We attribute this partly to the sharp edge structures in our monolayer (Figure S7, Supporting Information), arising from the relatively high growth temperature used for this work. To support this, Figure S8 (Supporting Information) shows the density-of-states of a local step edge of the monolayer, probed by STS, where finite electronic states are clearly evidenced around the Fermi level. This finding is in line with the DFT calculations for both armchair and zigzag edges of monolayer $VSe_2$ (Figure S9, Supporting Information). These edge-derived states not only can hide the CDW gap, but also induce additional energy broadening in the EDCs.

Our first experimental observation of spin frustration in monolayer $VSe_2$, a representative van der Waals monolayer magnet in the family of 2D-TMDs, is striking. As a frustrated magnet, its spins exhibit subtle correlations, albeit in the absence of a long-range order. Many exotic quantum phases, such as quantum spin liquid[46] and high-temperature superconductivity,[47] are related to spin frustration. We, therefore, expect this work to stimulate new work in the field of metallic 2D-TMDs in the theoretical and experimental exploration of quantum phenomena in the 2D limit. Moreover, our results suggest that atomic-scale structural details of monolayer $VSe_2$ may play a role in its magnetism, by modifying the Fermi level electronic structure and the relative stability of ferromagnetic and AF ground states. A systematic study establishing a structure-magnetism relationship will thus provide timely insights into this issue.

**Experimental Section**

*Molecular-Beam Epitaxy*: Monolayer VSe$_2$ films were grown on HOPG (SPI–1) in a custom-built MBE chamber with a base pressure of better than $1 \times 10^{-9}$ mbar. The substrates were prepared by *in-situ* cleavage followed by annealing at 820 K for at least 120 min. High-purity V and Se were evaporated from an electron-beam evaporator and a standard Knudsen cell, respectively. The Se/V flux ratio was controlled to be >10. During the growth process, the substrate temperature was kept at 650 K. To protect the samples from ambient contaminations during *ex-situ* transport to other UHV measurement systems, a Se capping layer with a thickness of at least ~10 nm was deposited on the samples after growth. For subsequent characterization by PES, ARPES, and XAS/XMCD, the Se cap was desorbed in UHV at 500 K for 20 minutes.

*Scanning Tunneling Microscopy and Spectroscopy*: STM measurements were carried out in a custom-built multi-chamber UHV system housing an Omicron LT-STM interfaced with a Nanonis controller. The base pressure was better than $10^{-10}$ mbar. A chemically etched tungsten tip was used, and the sample was kept at 77 K during all the measurements. STM images were recorded in constant current mode. For d$I$/d$V$ spectra, the tunneling current was obtained using the lock-in technique.

*Synchrotron-Based Photoemission*: PES measurements were performed at 300 K at the SINS beamline of the Singapore Synchrotron Light Source (SSLS), which covers the photon energy range from 50 to 1200 eV. A Scienta SES-200 spectrometer was used to collect the spectroscopic data at normal emission, while the X-ray beam was at an incident angle of 45º relative to the sample normal. A bias voltage of −7.0 V was applied to the sample during work function measurement in order to negate the effect of the analyzer work function. The binding energy of the data was calibrated using the 4$f$ core-levels and Fermi edge of a reference Au foil.

*Angle-Resolved Photoemission*: The ARPES measurements were collected with HeI$\alpha$ (h$\nu$=21.218 eV) radiation source (SCIENTA VUV5k). The photoelectrons were analyzed in the plane of incidence with a high energy and angular resolution SCIENTA DA30L analyzer. The angular detection range spans ±15° with respect to the spectrometer lens axes. Wider angular limits were obtained by rotating the sample with respect to the analyzer lens entrance axes. During ARPES acquisition the total energy resolution was set to 20 meV, with the angular resolution being better than 0.2°. The binding energy scale was referred to the Fermi level ($E_F$) as measured for a clean gold substrate.

*X-ray Absorption Spectroscopy and Magnetic Circular Dichroism*: XAS and XMCD measurements were independently carried out at the APE-HE beamline of Elettra Synchrotron Laboratory[48] and at the beamline 6.3.1 of the Advanced Light Source (ALS). The spectra were collected at a sample temperature range of 16–300 K in TEY mode, in which the sample drain current was recorded as a function of the photon energy. The angle of incidence of the photon beam was set to 45° relative to the sample normal. XMCD spectra were recorded with a fixed circular polarization of the X-rays and opposite magnetic fields up to ±1 T.

*Superconducting Quantum Interference Device*: Temperature- and magnetic field-dependent magnetic moments and susceptibility were characterized by Quantum Design SQUID over a temperature range of 2–300 K and fields up to 7 T.

*Calculations*: The electronic structure of monolayer VSe$_2$ was calculated within the density functional theory (DFT). The DFT calculations were performed using the VASP package, utilizing the projector augmented phase wave (PAW) method,[49] and the Perdew Burke and Ernzerhof (PBE) exchange-correlation functional.[50] An on-site Coulomb interaction was included within the GGA+U approach using the Dudarev method,[51] as implemented in VASP. We chose a value $U - J = 3$ eV for the V 3$d$ electrons. A separation of 15 Å between V layers in our supercell was found to be sufficient to represent an isolated monolayer. We employed a kinetic energy cutoff of 280 eV and a Γ-centered 20 × 20 × 1 k-point mesh. The lattice parameters and the atomic positions were optimized until the forces on the atoms were less than 10 meV/Å.

References

[1]     Q. H. Wang, K. Kalantar-Zadeh, A. Kis, J. N. Coleman, M. S. Strano, *Nat. Nanotechnol.* **2012**, *7*, 699.

[2]     M. Chhowalla, H. S. Shin, G. Eda, L. J. Li, K. P. Loh, H. Zhang, *Nat. Chem.* **2013**, *5*, 263.

[3]     D. Xiao, G. B. Liu, W. X. Feng, X. D. Xu, W. Yao, *Phys. Rev. Lett.* **2012**, *108*, 196802.

[4]     Y. Zhang, T. R. Chang, B. Zhou, Y. T. Cui, H. Yan, Z. K. Liu, F. Schmitt, J. Lee, R. Moore, Y. L. Chen, H. Lin, H. T. Jeng, S. K. Mo, Z. Hussain, A. Bansil, Z. X. Shen, *Nat. Nanotechnol.* **2014**, *9*, 111.

[5]     K. F. Mak, C. Lee, J. Hone, J. Shan, T. F. Heinz, *Phys. Rev. Lett.* **2010**, *105*, 136805.

[6]     K. F. Mak, K. L. He, J. Shan, T. F. Heinz, *Nat. Nanotechnol.* **2012**, *7*, 494.

[7]     H. L. Zeng, J. F. Dai, W. Yao, D. Xiao, X. D. Cui, *Nat. Nanotechnol.* **2012**, *7*, 490.

[8]     L. Y. Yang, N. A. Sinitsyn, W. B. Chen, J. T. Yuan, J. Zhang, J. Lou, S. A. Crooker, *Nat. Phys.* **2015**, *11*, 830.

[9]     X. X. Xi, L. Zhao, Z. F. Wang, H. Berger, L. Forro, J. Shan, K. F. Mak, *Nat. Nanotechnol.* **2015**, *10*, 765.

[10]    Y. J. Yu, F. Y. Yang, X. F. Lu, Y. J. Yan, Y. H. Cho, L. G. Ma, X. H. Niu, S. Kim, Y. W. Son, D. L. Feng, S. Y. Li, S. W. Cheong, X. H. Chen, Y. B. Zhang, *Nat. Nanotechnol.* **2015**, *10*, 270.

[11]    X. J. Zhu, Y. Q. Guo, H. Cheng, J. Dai, X. D. An, J. Y. Zhao, K. Z. Tian, S. Q. Wei, X. C. Zeng, C. Z. Wu, Y. Xie, *Nat. Commun.* **2016**, *7*, 11210.

[12]    L. J. Li, E. C. T. O'Farrell, K. P. Loh, G. Eda, B. Ozyilmaz, A. H. C. Neto, *Nature* **2016**,

*529*, 185.

[13]   A. K. Geim, I. V. Grigorieva, *Nature* **2013**, *499*, 419.

[14]   W. Q. Liu, P. K. J. Wong, Y. B. Xu, *Prog. Mater. Sci.* **2019**, *99*, 27.

[15]   J. Feng, X. Sun, C. Z. Wu, L. L. Peng, C. W. Lin, S. L. Hu, J. L. Yang, Y. Xie, *J. Am. Chem. Soc.* **2011**, *133*, 17832.

[16]   J. H. Zhou, L. Wang, M. Y. Yang, J. H. Wu, F. J. Chen, W. J. Huang, N. Han, H. L. Ye, F. P. Zhao, Y. Y. Li, Y. G. Li, *Adv. Mater.* **2017**, *29*, 1702061.

[17]   J. Feng, L. L. Peng, C. Z. Wu, X. Sun, S. L. Hu, C. W. Lin, J. Dai, J. L. Yang, Y. Xie, *Adv. Mater.* **2012**, *24*, 1969.

[18]   M. Y. Yan, X. L. Pan, P. Y. Wang, F. Chen, L. He, G. P. Jiang, J. H. Wang, J. Z. Liu, X. Xu, X. B. Liao, J. H. Yang, L. Q. Mai, *Nano Lett.* **2017**, *17*, 4109.

[19]   J. T. Yuan, J. J. Wu, W. J. Hardy, P. Loya, M. Lou, Y. C. Yang, S. Najmaei, M. L. Jiang, F. Qin, K. Keyshar, H. Ji, W. L. Gao, J. M. Bao, J. Kono, D. Natelson, P. M. Ajayan, J. Lou, *Adv. Mater.* **2015**, *27*, 5605.

[20]   K. Tsutsumi, *Phys. Rev. B* **1982**, *26*, 5756.

[21]   Y. D. Ma, Y. Dai, M. Guo, C. W. Niu, Y. T. Zhu, B. B. Huang, *ACS Nano* **2012**, *6*, 1695.

[22]   H. Pan, *J. Phys. Chem. C* **2014**, *118*, 13248.

[23]   M. Bonilla, S. Kolekar, Y. Ma, H. C. Diaz, V. Kalappattil, R. Das, T. Eggers, H. R. Gutierrez, P. Manh-Huong, M. Batzill, *Nat. Nanotechnol.* **2018**, *13*, 289.

[24]   G. Duvjir, B. K. Choi, I. Jang, S. Ulstrup, S. Kang, T. T. Ly, S. Kim, Y. H. Choi, C. Jozwiak, A. Bostwick, E. Rotenberg, J. G. Park, R. Sankar, K. S. Kim, J. Kim, Y. J. Chang, *Nano Lett.* **2018**, *18*, 5432.

[25]   J. G. Feng, D. Biswas, A. Rajan, M. D. Watson, F. Mazzola, O. J. Clark, K. Underwood, I. Markovic, M. McLaren, A. Hunter, D. M. Burn, L. B. Duffy, S. Barua, G. Balakrishnan, F. Bertran, P. Le Fevre, T. K. Kim, G. van der Laan, T. Hesjedal, P. Wahl, P. D. C. King, *Nano Lett.* **2018**, *18*, 4493.

[26]   P. Chen, W. W. Pai, Y. H. Chan, V. Madhavan, M. Y. Chou, S. K. Mo, A. V. Fedorov, T. C. Chiang, *Phys. Rev. Lett.* **2018**, *121*, 196402.

[27]   K. Xu, P. Z. Chen, X. L. Li, C. Z. Wu, Y. Q. Guo, J. Y. Zhao, X. J. Wu, Y. Xie, *Angew. Chem. Int. Ed.* **2013**, *52*, 10477.

[28]   A. Wasey, S. Chakrabarty, G. P. Das, *J. Appl. Phys.* **2015**, *117*, 064313.

[29]   D. J. O'Hara, T. Zhu, A. H. Trout, A. S. Ahmed, Y. K. Luo, C. H. Lee, M. R. Brenner, S. Rajan, J. A. Gupta, D. W. McComb, R. K. Kawakami, *Nano Lett.* **2018**, *18*, 3125.

[30]   T. Dietl, *Nat. Mater.* **2010**, *9*, 965.

[31]   Z. P. Zhang, J. J. Niu, P. F. Yang, Y. Gong, Q. Q. Ji, J. P. Shi, Q. Y. Fang, S. L. Jiang, H. Li, X. B. Zhou, L. Gu, X. S. Wu, Y. F. Zhang, *Adv. Mater.* **2017**, *29*, 1702359.

[32]   Y. Wang, Z. Sofer, J. Luxa, M. Pumera, *Adv. Mater. Interfaces* **2016**, *3*, 1600433.

[33]   M. Mulazzi, A. Chainani, N. Katayama, R. Eguchi, M. Matsunami, H. Ohashi, Y. Senba, M. Nohara, M. Uchida, H. Takagi, S. Shin, *Phys. Rev. B* **2010**, *82*, 075130.

[34]   D. Q. Gao, Q. X. Xue, X. Z. Mao, W. X. Wang, Q. Xu, D. S. Xue, *J. Mater. Chem. C* **2013**, *1*, 5909.

[35]   Z. L. Liu, X. Wu, Y. Shao, J. Qi, Y. Cao, L. Huang, C. Liu, J. O. Wang, Q. Zheng, Z. L. Zhu, K. Ibrahim, Y. L. Wang, H. J. Gao, *Sci. Bull.* **2018**, *63*, 419.

[36]   M. Farmanbar, G. Brocks, *Adv. Electron. Mater.* **2016**, *2*, 1500405.

[37]   G. Silversmit, D. Depla, H. Poelman, G. B. Marin, R. De Gryse, *J. Electron Spectrosc. Relat. Phenom.* **2004**, *135*, 167.

[38]   R. Claessen, I. Schafer, M. Skibowski, *J. Phys.-Condes. Matter* **1990**, *2*, 10045.

[39]   P. K. J. Wong, E. van Geijn, W. Zhang, A. A. Starikov, T. L. A. Tran, J. G. M. Sanderink, M. H. Siekman, G. Brocks, P. J. Kelly, W. G. van der Wiel, M. P. de Jong, *Adv. Funct. Mater.* **2013**, *23*, 4933.

[40]   P. K. J. Wong, W. Zhang, K. Wang, G. van der Laan, Y. B. Xu, W. G. van der Wiel, M. P. de Jong, *J. Mater. Chem. C* **2013**, *1*, 1197.

[41] N. B. Aetukuri, A. X. Gray, M. Drouard, M. Cossale, L. Gao, A. H. Reid, R. Kukreja, H. Ohldag, C. A. Jenkins, E. Arenholz, K. P. Roche, H. A. Durr, M. G. Samant, S. S. P. Parkin, *Nat. Phys.* **2013**, *9*, 661.

[42] Y. Zhou, K. Kanoda, T.-K. Ng, *Rev. Mod. Phys.* **2017**, *89*, 025003.

[43] L. Balents, *Nature* **2010**, *464*, 199.

[44] P. Fazekas, P. W. Anderson, *Phil. Mag.* **1974**, *30*, 423.

[45] N. Elstner, R. R. P. Singh, A. P. Young, *Phys. Rev. Lett.* **1993**, *71*, 1629.

[46] L. Clark, G. Sala, D. D. Maharaj, M. B. Stone, K. S. Knight, M. T. F. Telling, X. Wang, X. Xu, J. Kim, Y. Li, S.-W. Cheong, B. D. Gaulin, *Nat. Phys.* **2019**. DOI: 10.1038/s41567-018-0407-2

[47] P. A. Lee, *Rep. Prog. Phys.* **2008**, *71*, 012501.

[48] G. Panaccione, I. Vobornik, J. Fujii, D. Krizmancic, E. Annese, L. Giovanelli, F. Maccherozzi, F. Salvador, A. De Luisa, D. Benedetti, A. Gruden, P. Bertoch, F. Polack, D. Cocco, G. Sostero, B. Diviacco, M. Hochstrasser, U. Maier, D. Pescia, C. H. Back, T. Greber, J. Osterwalder, M. Galaktionov, M. Sancrotti, G. Rossi, *Rev. Sci. Instrum.* **2009**, *80*, 043105.

[49] G. Kresse, J. Furthmuller, *Comput. Mater. Sci.* **1996**, *6*, 15.

[50] J. P. Perdew, K. Burke, M. Ernzerhof, *Phys. Rev. Lett.* **1996**, *77*, 3865.

[51] S. L. Dudarev, G. A. Botton, S. Y. Savrasov, C. J. Humphreys, A. P. Sutton, *Phys. Rev. B* **1998**, *57*, 1505.

**Figure 1.** Crystal structure and STM/STS data of monolayer VSe$_2$. a) Crystal structure of 1$T$-VSe$_2$. b,c) Large-scale (50 × 50 nm$^2$; $V_{tip}$ = +1.0 V, $I_{set}$ = 200 pA) and atomic resolution (5 × 10 nm$^2$; $V_{tip}$ = +0.2 V, $I_{set}$ = 200 pA) STM images of monolayer VSe$_2$, measured at 77 K. The line profile in (b) shows a monolayer step height of ~7 Å, and that in (d) reveals an in-plane lattice parameter of 3.4 ± 0.1 Å. Both values are in good agreement with those in (a) for 1$T$-VSe$_2$. **e**, Averaged STS spectra of monolayer VSe$_2$ (set-point: $V_{tip}$ = +0.1 V, $I_{set}$ = 173 pA, 625 Hz, 50 mV) shows a differential conductance dip around the Fermi level with a size of ~26 meV, due to a CDW order. Its inset shows a zoom-in of the gap feature.

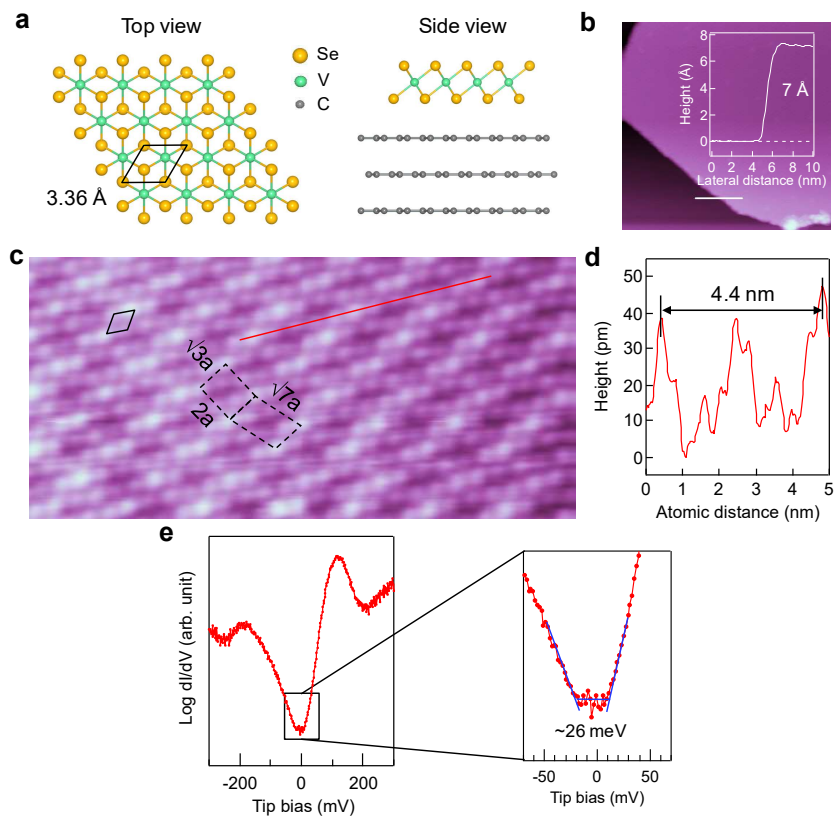

**Figure 2.** Synchrotron-PES data of pristine and ambient-exposed monolayer VSe$_2$. a,b) V $2p$ and Se $3d$ core-levels of monolayer VSe$_2$. The V $2p_{3/2}$ and $2p_{1/2}$ peaks are located at 513.4 eV and 520.9 eV, respectively, thus indicating a spin-orbit splitting of 7.5 eV. The Se $3d_{5/2}$ and $3d_{3/2}$ peaks are positioned at 53.4 eV and 54.2 eV, respectively. c) The valence band of the monolayer shows a sharp Fermi edge originating from the V $3d_z^2$ band. The spectral features between 1 and 5 eV are contributed by the Se $4p$ derived bands, while the broad bump beyond 6 eV is a substrate peak. d) Work function of monolayer VSe$_2$ extracted from the secondary electron cutoff, using a photon energy of 60 eV. The upper panel (red spectra) shows the effects of air-exposure. In particular, additional peaks, which are located at the higher BE side of the main peaks, are seen to develop at the V $2p$ and Se $3d$ core-levels. The sharp Fermi edge of the initial valence band has been reduced substantially, accompanied with a work function decrease of 0.25 eV.

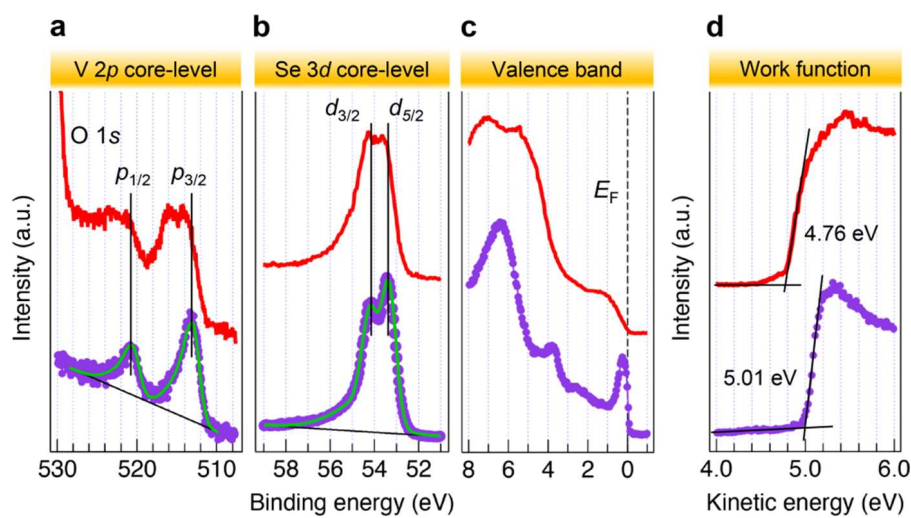

**Figure 3.** V $3d^1$ electronic configuration of 1T-VSe$_2$, and element-specific XAS and XMCD measurements of monolayer VSe$_2$. a) Schematic diagram of the $3d$ electronic states of 1T-VSe$_2$ in an octahedral crystal-field. The $3d$-orbitals of the V$^{4+}$ ion are split into two sets of $e_g$ and $t_{2g}$ orbitals, separated by an energy $\Delta_{oct} = 10\ Dq$. b) TEY detection geometry for the XAS and XMCD data shown in (c). c) Upper panel shows the XAS spectra of the monolayer measured at 30 K. Spectra highlighted in red and green are, respectively, the XAS obtained with opposite magnetic field (~±300 Oe) directions, and their sum gives the total XAS marked in purple. Atomic multiplet structure, identified as a-e, provides direct proof of the $d^1$ configuration. Lower panel in (c) shows the XMCD signal that is within the experimental error, indicating no signs of a ferromagnetic coupling in the monolayer. This result is supported by measurements taken at 16 K and with a 1 T field (Figure S5, Supporting Information).

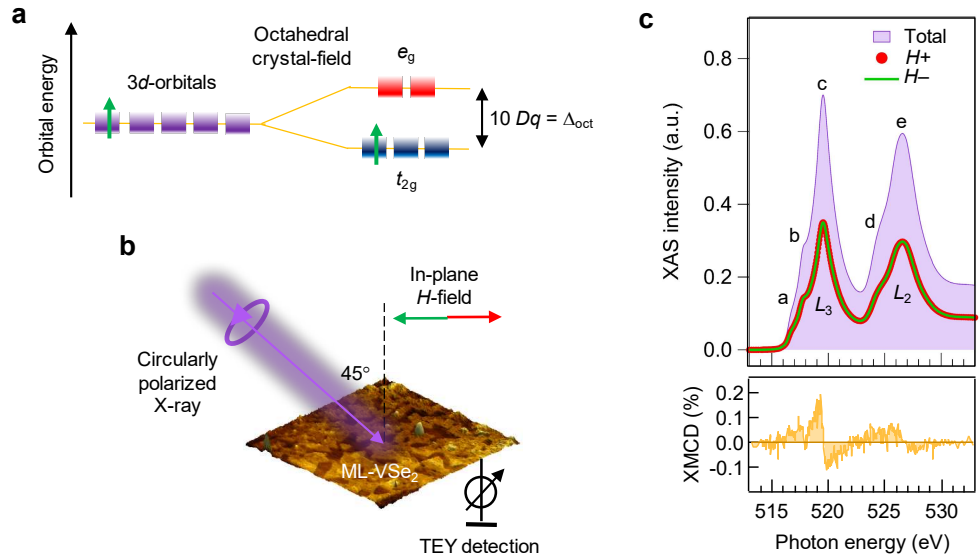

**Figure 4.** Temperature- and magnetic-field-dependent magnetic moment per area ($M$) and susceptibility ($\chi = M/H$) of monolayer VSe$_2$. a) In-plane and out-of-plane $M$–$H$ curves measured at 300 K. b) Out-of-plane $M$–$H$ curve as a function of temperature. c) upper panel, Magnetic susceptibility $\chi$ of monolayer VSe$_2$. An overlap between the field-cooling (FC) and zero-field-cooling (ZFC) curves down to 2 K and in a 7 T field, together with a broad maximum for the measured $\chi$, a characteristic feature of low-dimensional magnetic systems with a short-range antiferromagnetic (AF) interaction, indicates the presence of spin frustration in monolayer VSe$_2$. Lower panel, a negative Weiss constant $\theta_{cw}$ obtained from the Curie-Weiss (CW) fit to the high-temperature region of $1/\chi$, suggesting an antiferromagnetic (AF) origin of the exchange interaction in monolayer VSe$_2$. d) A triangular spin-lattice with AF coupling. In this geometry, three neighboring spins cannot be mutually antialigned, thus leading to spin fluctuation and suppressed AF correlation.

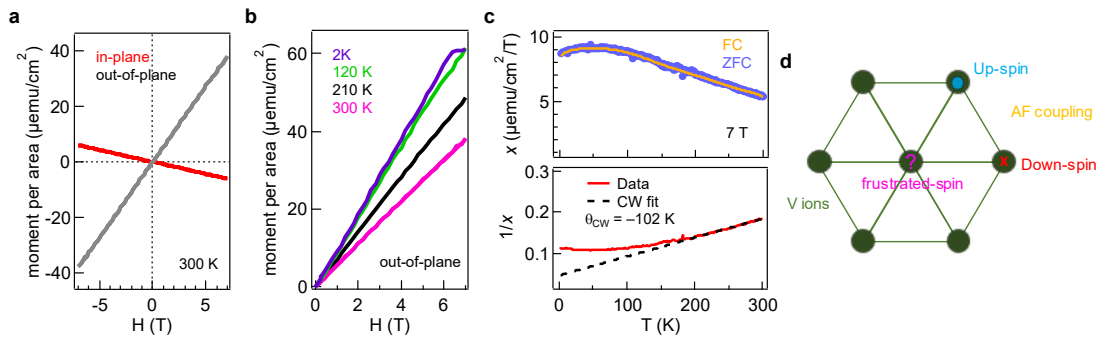

**Figure 5.** ARPES data of monolayer VSe$_2$. a) Intensity map as a function of the surface momentum component $k_{//}$ as acquired at 300 K. b) Zoom-in of (a) in the Fermi level BE range. c) Experimental band dispersion extracted from (b) using a second derivative filter. b-f) Same as (a-c), with data acquired at 11 K. Zero energy represents the Fermi level position. Upon cooling, the V 3$d$ and Se 4$p$ bands are observed to shift towards higher BE. g) These bands appear as a double-peak structure in the EDCs, which is not resolved at 300 K. The measured intensities of the EDCs have been normalized by the FD function, in order to highlight the effect of temperature on the spectral broadening. A CDW gap opening is evidenced by a leading-edge midpoint shift toward higher BE when cooling down to 11 K from 300 K.

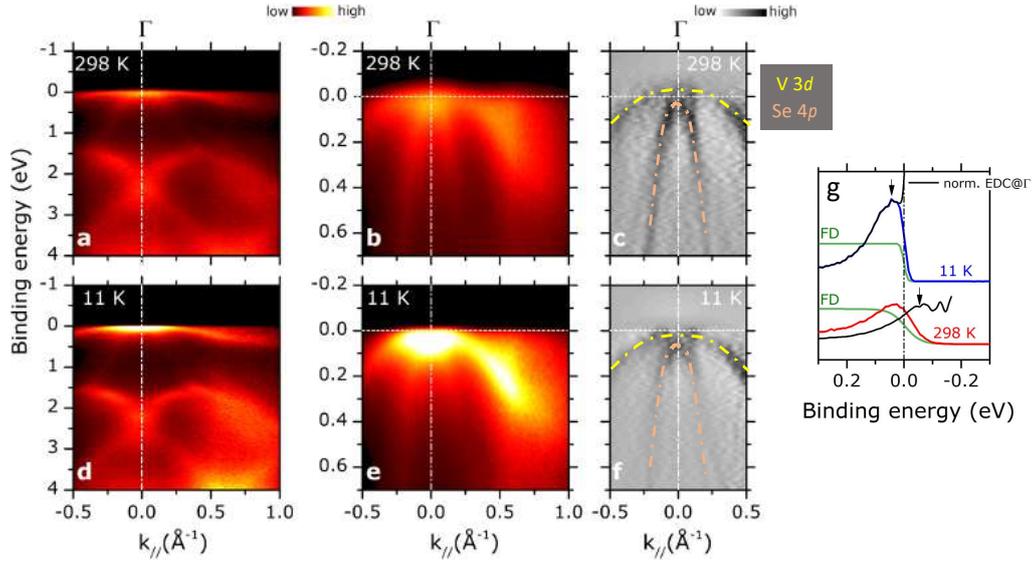

Supporting Information

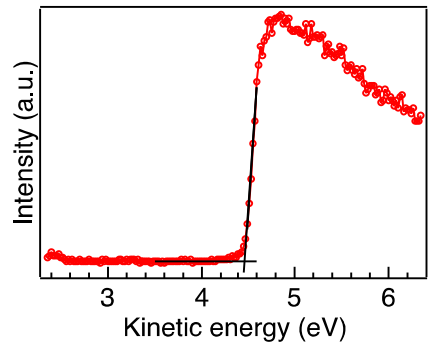

**Figure S1**. Work function measurement of bare HOPG, using a photon energy of 60 eV and a sample bias of –7.0 V. The work function extracted (4.46 ± 0.10 eV) is comparable with the known value for HOPG[1], and provides a calibration for monolayer VSe$_2$, as discussed in Figure 2d of the main text.

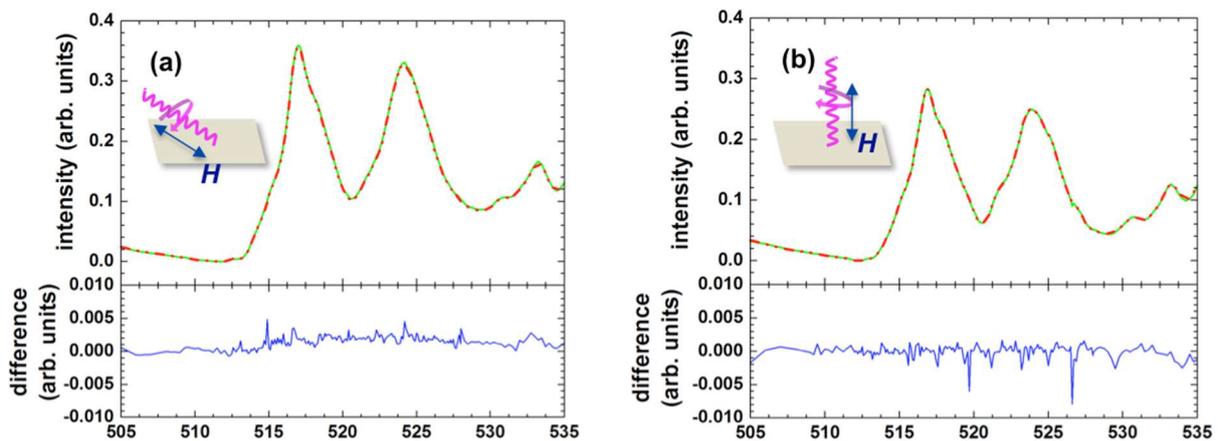

**Figure S2**. XMCD measurements of monolayer VSe$_2$ taken with 1 T magnetic fields at 16 K. The left panel shows the data obtained by a glancing incidence geometry, whereas the data in the right panel with a normal incidence geometry. No XMCD is observed for both geometries, indicating an absence of a FM coupling in the monolayer. The spectral features at 530–535 eV photon energy in both panels are attributed to adsorption of oxygen-containing molecules on the sample during the cryogenic measurements. These molecules, however, do not react with the sample and can be desorbed by warming up to 300 K.

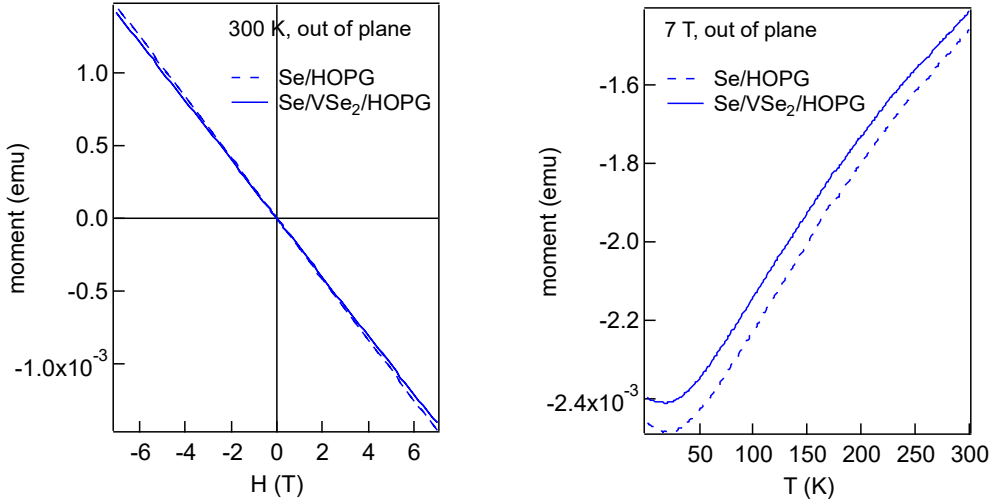

**Figure S3**. Out-of-plane magnetic moment measurements of Se/HOPG (dotted line) and Se/VSe$_2$/HOPG (solid line) at 300 K (left) and corresponding temperature-dependence obtained by SQUID with a magnetic field of 7 T (right).

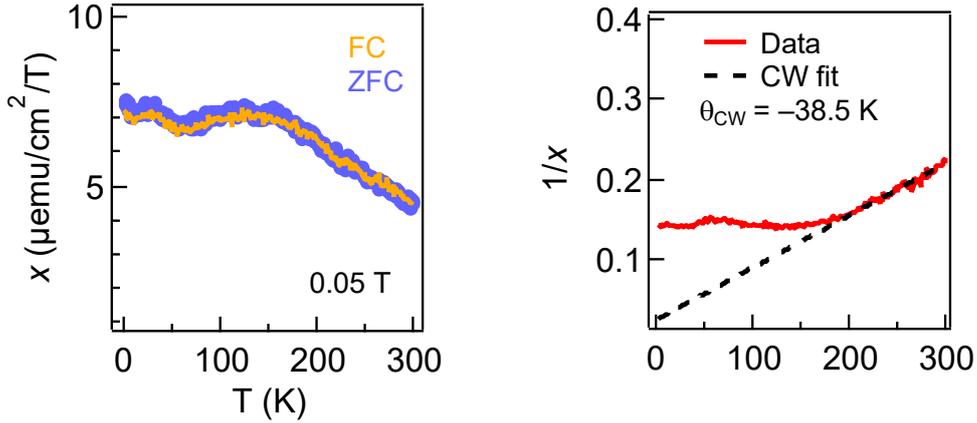

**Figure S4**. Temperature-dependent magnetic susceptibility $\chi$ (left) and its inverse, $\chi^{-1}$ (right), of monolayer VSe$_2$ obtained by SQUID with a magnetic field of 0.05 T. Similar to the 7 T data in **Figure 4**c, the FC and ZFC curves under 0.05 T (left) are found to overlap with each other. The temperature dependence of $\chi^{-1}$ is fit by Curie-Weiss law, with the formula $\chi = C/(T-\theta_{cw})$, where $C$ is the Curie constant and $\theta_{cw}$ is the Curie-Weiss temperature. Extraction of $\theta_{cw}$ provides a natural estimate for the strength of magnetic interactions, and is, in the present case, negative for monolayer VSe$_2$, indicating an AF coupling. We note that $\theta_{cw}$ is field-dependent, where an interplay between the AF interaction and CDW order in monolayer VSe$_2$ may play a role. With $H$ increasing from 0.05 T to 7 T, we find $|\theta_{cw}|$ to increase from ~38 K to 102 K, whose evolution is concomitant with a reducing CDW order from ~70 K to lower temperatures (Figure 4c). We tentatively correlate this observation with a progressively reduced CDW gap by the increasing $H$.

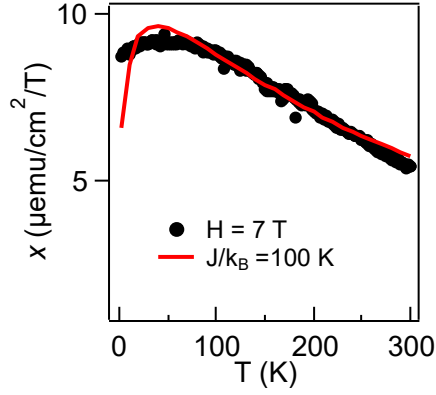

**Figure S5.** Quantitative fit to the $\chi$ of monolayer VSe$_2$ measured in a 7 T field, using the model of spin-1/2 Heisenberg antiferromagnet on a triangular lattice.[2] The black dots are the experimental data, and the red line represents the theoretical fit with $J/k_B$ = 100 K, where $J$ is the exchange coupling parameter. The reasonable reproducibility of the $\chi$ by this model suggests that frustration is operative in our monolayer VSe$_2$. Other models without considering spin frustration, such as that of spin-1/2 Heisenberg antiferromagnet on a 2D square lattice,[3, 4] fail to reproduce the broad peak in our experimental result. We also notice a divergence between the experimental and calculated $\chi$ at low temperatures, which may be related to the coexisting CDW order in the monolayer.

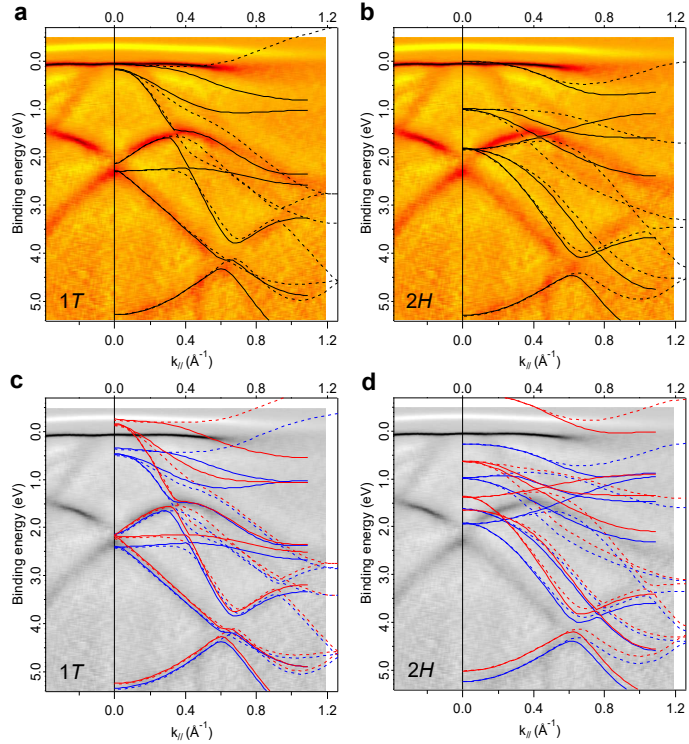

**Figure S6.** Overlays of the experimental ARPES dispersions (**Figure 5**) of monolayer VSe$_2$ with the calculated (**a,b**) non-spin-polarized and (**c,d**) spin-polarized band structures of both 1$T$ and 2$H$ phases, by DFT. Comparison between these bands confirms the 1$T$ structure of our monolayer. Moreover, a single $d$-band is observed in our ARPES data, thus excluding the theoretically predicted ferromagnetic ground state with an exchange splitting of more than 500 meV between the spin majority and minority bands near the Fermi level.[5] Note that the dispersions along the Γ–M direction in the hexagonal Brillouin zone are in solid lines and those along the Γ–K direction in dotted lines.

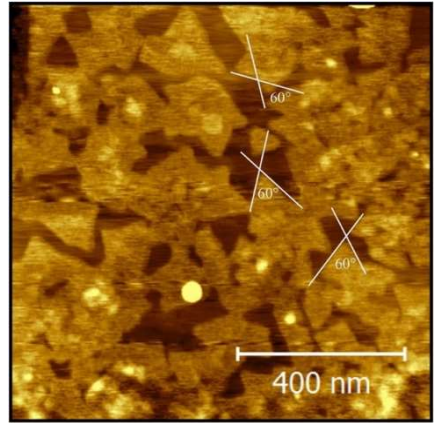

**Figure S7.** Representative *ex-situ* atomic force microscopy image of monolayer VSe$_2$ grown on HOPG. Without an amorphous Se cap, the monolayer is likely (partially) oxidized, causing a roughened surface. Nevertheless, following their sharp edges, one can still identify the highly-crystalline VSe$_2$ domains, as highlighted in the figure. Most of these domains are triangular in shape with an inner angle of 60°, in good agreement with VSe$_2$'s hexagonal atomic registry. However, relative misalignments of those domains within the scanned area (much smaller than the HOPG grain size of the order of 1 mm) suggest rotational disorder in the monolayer, arising from the rather weak van der Waals-type film-substrate interaction, which is also known for other MBE-grown 2D-TMD systems.[6, 7]

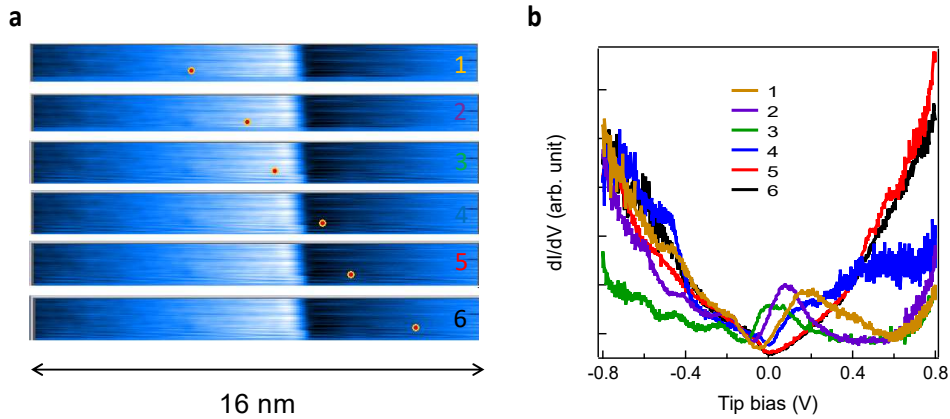

**Figure S8.** A series of STS measurements across the step edge of monolayer VSe$_2$ on HOPG at 77 K. Since all the data is obtained at the same measurement conditions (set-point: $V_{\text{tip}} = -0.3$ V, $I_{\text{set}} = 60$ pA, 625 Hz, 50 mV) and the HOPG substrate does not exhibit any sharp feature near the Fermi level, we are able to follow an evolution of the local density-of-states when moving from the VSe$_2$ terrace (point 1), its step edge (point 2,3), to HOPG (point 4-6). Essentially, we observe that the differential conductance at the Fermi level progressively increases when the tip moves closer to the edge. This indicates edge-related states emerged around the Fermi level, and might serve to explain the finite Fermi-edge cut-off in our ARPES measurements (Figure 5), despite the presence of a CDW order at low temperatures.

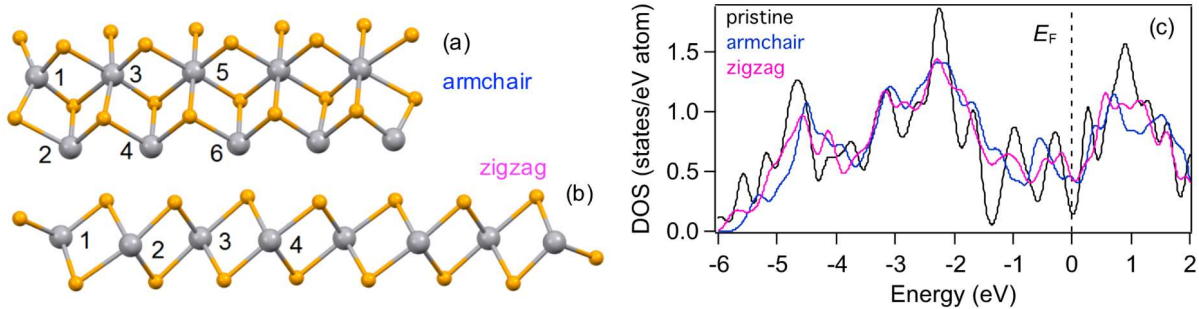

**Figure S9**. DFT calculations of the electronic structure of VSe$_2$ armchair (a) and zigzag (b) nanoribbons. Compared with the case of pristine VSe$_2$, both structures create an enhanced DOS at $E_F$, consistent with the STS measurements in Figure S8. For the modeling, self-consistent optimization of lattice parameter and atomic positions was performed using the Quantum ESPRESSSO code and the GGA-PBE. The energy cutoffs were 25 and 400 Ry for the plane-wave expansion of the wave functions and the charge density, respectively, and the 4×4×3 Monkhorst-Pack $k$-point grid for the Brillouin sampling. The number in (a,b) corresponds to non-equivalent vanadium atoms.

**Table S1.** Relative total energies (in meV per formula unit) of VSe$_2$ monolayers in the 1$T$ and 2$H$ phases, calculated by DFT using the PBE and the PBE+$U$ functionals (for details, see the Methods section in the main text). The FM state of the $T$-phase is used as reference. Our calculations indicate that the structures and magnetic ground states of monolayer VSe$_2$ are sensitive to the on-site Coulomb parameter $U$. Without $U$, the FM ground state of the $H$-phase is found to be more stable than the FM state of the $T$-phase by 88 meV/f.u., whereas the AF state of both $T$- and $H$-phases are higher in energy. Here we set the AF ordering by doubling the VSe$_2$ in-plane unit cell in one direction, and enforce an AF ordering of the magnetic moments on the two V atoms within this cell. Choosing a moderate value $U$ = 3 eV for the 3$d$ electrons of V renders the AF state of the $T$-phase more stable than the FM state of the $T$-phase by 22 meV/f.u., whereas both the FM and AF states of the $H$-phase are higher in energy. In general, the energy differences between the FM and AF states, and between the $T$- and $H$-phases tend to be small, and it is difficult to provide a conclusive remark on the basis of DFT calculations alone.

|  | Ferromagnetic (FM) (meV) | Antiferromagnetic (AF) (meV) |
| --- | --- | --- |
| *T-phase (PBE)* | 0 | 51 |
| *H-phase (PBE)* | –88 | 61 |
| *T-phase (+U)* | 0 | –22 |
| *H-phase (+U)* | 9 | 297 |

References

[1] H. Hibino, H. Kageshima, M. Kotsugi, F. Maeda, F. Z. Guo, Y. Watanabe, *Phys. Rev. B* **2009**, *79*, 125437.

[2] N. Elstner, R. R. P. Singh, A. P. Young, *Phys. Rev. Lett.* **1993**, *71*, 1629.

[3] J. Wang, *Phys. Rev. B* **1991**, *44*, 2396.

[4] F. M. Woodward, A. S. Albrecht, C. M. Wynn, C. P. Landee, M. M. Turnbull, *Phys. Rev. B* **2002**, *65*, 144412.

[5] Y. D. Ma, Y. Dai, M. Guo, C. W. Niu, J. B. Lu, B. B. Huang, *Phys. Chem. Chem. Phys.* **2011**, *13*, 15546.

[6] M. M. Ugeda, A. J. Bradley, S. F. Shi, F. H. da Jornada, Y. Zhang, D. Y. Qiu, W. Ruan, S. K. Mo, Z. Hussain, Z. X. Shen, F. Wang, S. G. Louie, M. F. Crommie, *Nat. Mater.* **2014**, *13*, 1091.

[7] M. M. Ugeda, A. J. Bradley, Y. Zhang, S. Onishi, Y. Chen, W. Ruan, C. Ojeda-Aristizabal, H. Ryu, M. T. Edmonds, H. Z. Tsai, A. Riss, S. K. Mo, D. H. Lee, A. Zettl, Z. Hussain, Z. X. Shen, M. F. Crommie, *Nat. Phys.* **2016**, *12*, 92.